\documentclass[journal]{IEEEtran}
\usepackage{amsmath}    
\usepackage{graphics,epsfig,subfigure}    
\usepackage{verbatim}   
\usepackage{color}      

\usepackage{subfigure}  

\usepackage{amssymb}
\usepackage{epstopdf}
\usepackage{graphicx}
\graphicspath{{/Figures/}}
\usepackage{wasysym}
\usepackage{wrapfig}
\usepackage{sidecap}
\usepackage{float}
\usepackage{subfigure}
\usepackage{enumerate}
\usepackage{graphicx}
\graphicspath{{./figures/}}
\usepackage{algorithmic}
\usepackage{algorithm}
\usepackage{amsthm}
\usepackage{bookmark}
\usepackage{algorithmic}
\usepackage{algorithm}
\usepackage{amsthm}
\usepackage{multirow}
\usepackage{float}
\floatstyle{ruled}
\newfloat{algorithm}{htbp}{loa}
\floatname{algorithm}{Algorithm}

\newtheorem{problem}{Problem}

\begin{document}
\title{Coordinated Autonomous Vehicle Parking for Vehicle-to-Grid Services: Formulation and Distributed Algorithm}

\author{Albert Y.S. Lam, James J.Q. Yu, Yunhe Hou, and
        Victor O.K. Li
\thanks{This work was supported by the Theme-Based Research Scheme of the Research Grants Council of Hong Kong under Grant T23-701/14-N. A preliminary version of this paper was presented in \cite{smartgridcomm2016}. The authors are with the Department
of Electric and Electronic Engineering, The University of Hong Kong, Pokfulam Road, 
Hong Kong (e-mail: ayslam@eee.hku.hk).}
}


\maketitle

\begin{abstract}
Autonomous vehicles (AVs) will revolutionarize ground transport and take a substantial role in the future transportation system. Most AVs are likely to be electric vehicles (EVs) and they can participate in the vehicle-to-grid (V2G) system to support various V2G services. Although it is generally infeasible for EVs to dictate their routes, we can design AV travel plans to fulfill certain system-wide objectives. In this paper, we focus on the AVs looking for parking and study how they can be led to appropriate parking facilities to support V2G services. We formulate the Coordinated Parking Problem (CPP), which can be solved by a standard integer linear program solver but requires long computational time. To make it more practical, we develop a distributed algorithm to address CPP based on dual decomposition. We carry out a series of simulations to evaluate the proposed solution methods. Our results show that the distributed algorithm can produce nearly optimal solutions with substantially less computational time.  A coarser time scale can improve computational time but degrade the solution quality resulting in possible infeasible solution. Even with communication loss, the distributed algorithm can still perform well and converge with only little degradation in speed.

\end{abstract}

\IEEEpeerreviewmaketitle

\textcolor{black}{
\section*{Nomenclature}
\addcontentsline{toc}{section}{Nomenclature}
\begin{IEEEdescription}[\IEEEusemathlabelsep\IEEEsetlabelwidth{$V_1,V_2,V_3$}]
\item[$G$] The complete directed graph modeling the road network.
\item[$\mathcal{N}$] Set of all possible locations.
\item[$\mathcal{E}$] Set of paths connecting the locations.
\item[$d_{ij}$] Expected travel distance from $i$ to $j$.
\item[$\mathcal{K}$] Set of autonomous vehicles (AVs).
\item[$k$] A particular AV.
\item[$\underline{n}_k$] Initial location of AV $k$.
\item[$\overline{n}_k$] Return location of AV $k$.
\item[$\underline{t}_k$] Initial time of AV $k$ available for parking.
\item[$\overline{t}_k$] End time of AV $k$ for parking.
\item[$\underline{e}_k$] State of charge (SOC) of AV $k$ right before parking.
\item[$\overline{e}_k$] Expected SOC of AV $k$ on return.
\item[$d_k^{max}$] Maximum distance AV $k$ allowed to travel.
\item[$\alpha_k$] Estimation function of AV $k$.
\item[$f$] A particular parking facility.
\item[$\hat{n}_f$] Location of parking facility $f$.
\item[$\underline{m}_k$] Duration for AV $k$ to reach the parking facility from its initial location.
\item[$\overline{m}_k$] Duration for AV $k$ to reach the return location from the parking facility.
\item[$\underline{\epsilon}_k$] Amount of energy required for AV $k$ to reach the parking facility from its initial location.
\item[$\overline{\epsilon}_k$] Amount of energy required for AV $k$ to reach the return location from the parking facility.
\item[$\mathcal{F}$] Set of parking facilities.
\item[$p_f$] Demand profile of parking facility $f$.
\item[$t$] A particular time slot.
\item[$\rho_t^f$] Number of AVs required to support the services at parking facility $f$ in the $t$th time slot.
\item[$D$] Latest time slot in the time horizon.
\item[$c_f$] Capacity of parking facility $f$.
\item[$\beta_f$] Estimation function of parking facility $f$.
\item[$\hat{m}_k^f$] Duration that AV $k$ should stay at parking facility $f$.
\item[$e'_k$] SOC of AV $k$ when arriving at parking facility $f$.
\item[$e''_k$] SOC of AV $k$ when leaving parking facility $f$.
\item[$\mathcal{T}$] Time horizon.
\item[$x_{kt}^{f}$] Binary variable to indicate if AV $k$ is assigned to parking facility $f$ at the time slot $t$.
\item[$y_{k}^{f}$] Binary variable to indicate if AV $k$ is parked at parking facility $f$.
\item[$M$] A sufficiently large positive number.
\item[$\underline{\lambda}_t^f$] Lagrangian multiplier corresponding to $\rho_t^f$.
\item[$\overline{\lambda}_t^f$] Lagrangian multiplier corresponding to $c_f$.
\item[$\underline{\Lambda}$] Vector of Lagrangian multiplier $\underline{\lambda}_t^f$.
\item[$\overline{\Lambda}$] Vector of Lagrangian multiplier $\overline{\lambda}_t^f$.
\item[$g(\overline{\Lambda},\underline{\Lambda})$] Dual function.
\item[$\underline{\gamma}_t^f(i)$] Step size for the update rule of $\underline{\lambda}_t^f$ at Iteration $i$.
\item[$\overline{\gamma}_t^f(i)$] Step size for the update rule of $\overline{\lambda}_t^f$ at Iteration $i$.
\item[$x_{kt}^{f*}(i)$] Optimal result by solving Problem 2 at Iteration $i$.
\item[$\delta$] A small positive number.
\item[$\gamma^{cap}(i)$] Cap of step size at Iteration $i$.
\item[$\gamma^{init}$] Initial value of the step size.
\item[$\epsilon$] A small positive number.
\end{IEEEdescription}
}

\section{Introduction}

\IEEEPARstart{T}{hanks}  
 to people's stronger environmental awareness and various governments' green policies, increasingly more electric vehicles (EVs) will run on the roads. EVs largely rely on the grid to charge their batteries. Besides, they can also discharge any excessive energy back to the grid. The EV batteries become a significant yet flexible energy repository. This vehicle-to-grid (V2G) system which can complement the grid with various demand response and auxiliary services. A V2G system \textcolor{black}{may be considered to be} associated with a parking facility where a large number of EVs can contribute their batteries to support various V2G services \textcolor{black}{\cite{V2G_capacity}}.
\textcolor{black}{However, convenience plays a very important role when an EV driver decides where and when to park its vehicle. EV mobility behavior is considered stochastic \cite{EVmobility} and it is hard for a parking facility to predict accurately how many EVs will be available in a particular period, even in the next few hours.}

Autonomous vehicles (AVs), also known as driverless cars and robotic cars, refer to those vehicles which can navigate without human intervention. Based on the recent trend of the automotive industry, e.g., from Tesla, AVs will become prevalent on the roads. \textcolor{black}{It has been predicted that AVs will  revolutionize the automobile industry in the next two decades \cite{trend1, trend2, trend3}.} They are equipped with numerous sensors to facilitate their interactions with the surrounding environments. An AV may be fully or partially driverless; a driver can guide the movement in the ``normal'' mode and it can also implement self-navigation in the ``autonomous'' mode without the driver's input.
AVs enjoy many advantages over conventional cars, like avoiding collisions due to human errors, lessening traffic congestion, and reducing physical space for vehicle parking.

AVs are typically electric and they contain batteries to store energy for propulsion. Hence, AVs can participate in V2G. Due to their self-driving ability and advanced vehicular communication technologies, AVs can be coordinated to orchestrate more co-operative exercises.
\textcolor{black}{It is possible to arrange an appropriate number of AVs with parking intention to the right location to support V2G services. Hence, }
AVs are \textcolor{black}{considered} advantageous over ordinary EVs in the sense that the intrinsic uncontrollable EV behaviors, with respect to their appearance at V2G infrastructure, can now be overcome. Moreover, different V2G-supporting parking facilities have diverse V2G objectives and they have different ``demands'' of EVs anchoring at the facilities at different times. We can now deploy more effective V2G services by appropriately assigning AVs to the parking facilities to meet their EV demands. Therefore, in this paper, instead of studying how AVs contribute to V2G in parking facilities directly, we investigate how to coordinate AV parking to facilitate V2G services. To the best of our knowledge, we are the first to study how to manage AVs for supporting V2G services. 
We formulate the Coordinated Parking Problem (CPP) for AVs to support V2G. While a centralized and a heuristic solution have been proposed in our preliminary version \cite{smartgridcomm2016}, we develop a distributed algorithm to make the problem solving scalable so that this work can become more practical.  Compared with  \cite{smartgridcomm2016}, our contributions include: (1) conducting a more comprehensive literature review; (2) providing a neater formulation of the problem with fewer constraints; (3) developing an effective distributed algorithm to address the problem; and (4) conducting extensive simulation to evaluate the performance of the distributed algorithm and to compare with the centralized and heuristic approaches proposed in \cite{smartgridcomm2016}.

The rest of the paper is organized as follows. Section \ref{sec:related} provides the related work. We develop models for the road network, AVs, and parking facilities and illustrate the system operation in Section \ref{sec:model}. Section \ref{sec:formulation} formulates CPP as an optimization problem, and we develop an effective distributed algorithm in Section \ref{sec:distributed}. In Section \ref{sec:performance}, we evaluate the performance of the various solution methods and conclude the paper in Section \ref{sec:conclusion}.

\section{Related Work} \label{sec:related}
There are many related efforts studying the relationship between V2G and the supported services.
\cite{V2G_DR} investigated how demand response helps reduce peak power demand and shape the V2G aggregated demand profile. \cite{V2G_DR2} studied the impact of EV mobility on demand response for V2G and presented a dynamic complex network model of V2G mobile energy networks.  
In \cite{V2G_ancillary2}, an EV scheduling algorithm was developed to optimize bidding of V2G for various ancillary services, including frequency regulation and spinning reserve. It maximizes the aggregator's profit while providing peak load shaving to the utility. \cite{V2G_ancillary} formulated the optimal combined bidding of V2G ancillary services and it can enhance the profit of the aggregators, utilities, and EV customers.
 \cite{V2G_frequencyregulation} designed a V2G aggregator for frequency regulation and a dynamic programming algorithm was developed to control the optimal charging for the vehicles.
\cite{V2G_capacity} estimated the capacity of V2G for frequency regulation with a queueing network model. 
%
\textcolor{black}{\cite{V2GService1} discussed the economical operation of energy resources, like batteries, for microgrid.
\textcolor{black}{\cite{V2GService2} proposed a distributed EV coordination management for efficient exploitation of renewable energy.}
}
All these suggest that V2G may potentially be beneficial to the grid and one of the keys to success is to ensure the availability of EVs to participate in V2G.

There are numerous research projects related to AVs.
For instance, \cite{AV_obstacle} designed an obstacle avoidance motion control scheme for AVs operating in uncertain dynamic environments.
\cite{AV_path} developed a hierarchical controller for AVs to track reference paths in uncertain  conditions and with external disturbances. 
\cite{kinect} designed a method to detect obstacles and dangerous areas in the outdoor environments with Kinect sensors installed on AVs.
\cite{AV_flocking} studied the collective behavior of AV flocking under an all-to-all communication scheme.
In \cite{AVPTS}, AVs co-operated in a public transportation system, in which AVs were scheduled with centralized control. Admission control of the system was also fully investigated. \cite{AVPTS_auction} focused on the pricing issue of the AV public transportation system and developed a combinatorial auction-based strategy-proof pricing scheme. 
The automotive industry is also developing AV technologies. 
\cite{connectedAV} reported the state-of-the-art development in the AV industry and AVs will become connected vehicles. The success of AVs will rely on connectivity and cooperation of the vehicles.
Google launched the self-driving car project and built a fully functioning prototype without a steering wheel and pedals \cite{googlecar}. A Tesla car can enable its autonomous driving ability with a software update \cite{teslacar}. Thus AVs are not just idle theorizing and they can have practical use sooner or later.

There is some work studying intelligent parking in general.
\cite{parkingassistance} investigated availability of parking facilities for parking guidance and information systems. It developed a multivariate autoregressive model to account for the temporal and spatial relationship of parking availability.
\cite{streetparking} studied the uncoordinated parking space allocation for inexpensive limited on-street parking spots and expensive oversized parking lots.  
Some work focuses on AV parking.
\cite{valet1} developed a control system for AV valet parking with a focus on steering control.
\cite{valet2} designed an intelligent vehicle system to implement the AV valet parking service.
However, they mainly targeted AV parking control in a confined parking area.
Some investigate the parking issue for a larger area.
\cite{parkingmanagement} proposed intelligent parking assistant architecture to manage parking spots to improve the quality of urban mobility.
\cite{grid_parking} analyzed the impact of charging and discharging of EVs in parking lots on the power grid probabilistically.
However, there is no thorough study on V2G based on AVs. In this work, we aim to bridge this research gap.

\section{System Model} \label{sec:model}
The system is composed of three types of components, namely, a road network, AVs, and parking facilities. \textcolor{black}{In this section, we first describe the required infrastructure and then} provide models of these system components. \textcolor{black}{Finally we illustrate} how the system operates. 

\textcolor{black}{
\subsection{Infrastructure}
Consider that we implement the coordinated AV parking in a smart city \cite{smartcity}, in which the road infrastructure is well-established. There is full communication coverage, backed by advanced vehicular communication technologies (e.g., IEEE 802.11p), supporting various intelligent transportation systems applications. A control center, implemented in the cloud, acts as the central ``brain'' of the system to manage the fleet of AVs with parking intention and the parking facilities. The Internet of Things backbone provides real-time communication support between the AVs (parking facilities) and the control center. The control center collects the required information from the AVs and parking facilities, does the computation, and gives instructions to the AVs for parking arrangement. A similar infrastructure is also adopted in \cite{AVPTS} and \cite{AVPTS_auction} to implement an AV-based public transportation system.
}

\subsection{Road Network}
We describe the accessibility of the AVs to and from the parking facilities with a road network. The road network is modeled by a complete directed graph $G(\mathcal{N},\mathcal{E})$, where $\mathcal{N}$ is the set of all possible locations where the AVs and the parking facilities are located.  $\mathcal{E}$ represents the set of paths connecting the locations. Each $(i,j)\in \mathcal{E}$ is associated with the distance $d_{ij}$ indicating the expected travel distance from $i$ to $j$.  $d_{ij}$ is in general not equal to $d_{ji}$ and this accounts for the possible asymmetry of travel distances in different directions. Note that $G(\mathcal{N},\mathcal{E})$ is not a direct representation of the corresponding road network; in $G(\mathcal{N},\mathcal{E})$, a node is always accessible by another node in one hop. We can construct $G(\mathcal{N},\mathcal{E})$ from the road system by specifying a route from $i$ to $j$ with the corresponding distance, for each $(i,j)$ pair. For instance, we may employ Dijkstra's algorithm \cite{dijkstra} to suggest the shortest route to connect $i$ to $j$. We assume $d_{ij}$'s are static at the time of assignment. $d_{ij}$'s can be revised to reflect the updated traffic conditions in any subsequent assignments.

\subsection{Autonomous Vehicles}
\label{subsec:AV}
We denote the set of AVs which need parking by $\mathcal{K}$. Each $k\in\mathcal{K}$ is specified by the tuple $\langle \underline{n}_k,\overline{n}_k, \underline{t}_k,\overline{t}_k,\underline{e}_k,\overline{e}_k, d^\textit{max}_k, \alpha_k \rangle$. The autonomous parking mode of $k$ is turned on at $\underline{n}_k\in\mathcal{N}$ at time $\underline{t}_k$ with state of charge (SOC) $\underline{e}_k$ and it is expected to return to $\overline{n}_k\in\mathcal{N}$  by time $\overline{t}_k$ ($\overline{t}_k\geq \underline{t}_k$) with SOC  $\overline{e}_k$, which represents the minimum allowable SOC of the battery when the driver uses the car again after parking. $\underline{n}_k$  is allowed to be different from $\overline{n}_k$ for the convenience of the driver. As $k$ is expected to park in one of the parking facilities, the driver may desire to confine the total distance that the AV travels during $(\underline{t}_k,\overline{t}_k)$. The maximum distance that AV $k$ is allowed to travel in the autonomous mode is indicated by $d_k^\textit{max}$.\footnote{\textcolor{black}{$d_k^\textit{max}$ is not the maximum range supported by the energy stored in the battery of AV $k$. Instead, it is a value set by the owner who tries to cap the distance traversed for parking. This value is generally small and thus the range limit due to energy sufficiency does not matter.}} If the assigned parking facility $f$ is known, the AV can estimate the amount of time and energy required to reach $f$ from $\underline{n}_k$ and those required to arrive at $\overline{n}_k$ from $f$ based on the relevant details (including its locations, driving speed, and energy consumption rate). We define the function $\alpha_k$ to accomplish such estimation as
\begin{align}
[\underline{m}_k,\overline{m}_k,\underline{\epsilon}_k,\overline{\epsilon}_k]=\alpha_k(\underline{n}_k,\overline{n}_k,\underline{t}_k,\overline{t}_k,\hat{n}_f),
\label{AV2PF1}
\end{align}
where $\hat{n}_f$, $\underline{m}_k$, and $\overline{m}_k$   refer to the location of $f$, the duration for $k$ to reach $\hat{n}_f$ from $\underline{n}_k$ and the duration for $k$ to return to $\overline{n}_k$  from $\hat{n}_f$, respectively. $\underline{\epsilon}_k$ and $\overline{\epsilon}_k$ are the amounts of energy required to support the first and second legs of the parking journey, respectively. \textcolor{black}{Thus the reduced energy for mobility needs has been captured.}

\subsection{Parking Facilities}
\label{subsec:PF}
We consider a set of parking facilities $\mathcal{F}$, each of which represents a V2G system connected to the grid as in \cite{V2G_capacity}. Each $f\in\mathcal{F}$ is described by the tuple $\langle \hat{n}_f,p_f,c_f,\beta_f\rangle$.  $p_f=[\rho^f_t]_{1\leq t\leq D}$ denotes the demand profile of $f$, where $\rho_t^f$ gives the number of AVs required to support the V2G services at $f$ in the $t$th time slot and $D$ is the latest time slot in the time horizon (The time slot operation will be explained in Section \ref{subsec: operation}). There is much work in the literature describing how to utilize EVs to facilitate different kinds of V2G services, e.g., frequency regulation \cite{V2G_capacity}. The basic principle is that, for $f$ to provide various V2G services, it needs to acquire a certain number of vehicles for charging and discharging. Here we model the demand on the vehicles for V2G for the given time horizon by $p_f$. $c_f$ denotes the capacity of $f$ dedicated to the current operation. In other words, it represents the number of AVs which $f$ can accommodate in the time horizon.  We assume that $f$ is capable of determining how long AV $k$ should park at $f$. In this parking duration, $k$ will be charged up to a level that at least $\overline{e}_k$ will be retained when reaching $\overline{n}_k$, with the consideration of an appropriate charging rate and the amount of energy charged or discharged to support V2G. Consider that $f$ can facilitate the estimation with the function $\beta_f$ based on the SOC specifications  of AV $k$ as
\begin{align}
\hat{m}_k^f=\beta_f(e'_k,e''_k),
\label{AV2PF2}
\end{align}
where $\hat{m}_k^f$ is the duration that $k$ should stay at $f$. $e_k'=\underline{e}_k-\underline{\epsilon}_k$ and $e_k''=\overline{e}_k+\overline{\epsilon}_k$ represent the SOCs of $k$ when arriving at $f$ and when leaving from $f$, respectively, where $\underline{\epsilon}_k$ and $\overline{\epsilon}_k$ are computed from \eqref{AV2PF1}.
\textcolor{black}{
In other words, given the SOC requirements of AV $k$ in terms of $e'_k$ and $e''_k$, $f$ can manage the V2G events applied to $k$ (this may charge or discharge the battery of $k$) and determine an appropriate charging profile for $k$. When $k$ leaves $f$, $f$ will ensure $k$'s SOC reached  $e''_k$ by keeping $k$ at $f$ for $m_k^f$ time slots. In the literature, a lot of existing work (e.g., \cite{V2G_capacity, V2GApp1, V2GApp2, V2GApp3, V2GApp4}) has already investigated the energy management of vehicles and their interactions with the grid for V2G. In this work, we do not plan to replicate these efforts and simply represent all these by $\beta_f$. For a particular V2G application, we can construct the corresponding $\beta_f$ based on the relevant published work. In this way, we can simplify our model and pay our attention to AV parking arrangement, which is the main theme of this paper.
}

\subsection{Operation}
\label{subsec: operation}
Suppose that there is a control center which \textcolor{black}{co-ordinates} the parking of AVs. This control center aims to serve a dedicated group of AVs, e.g., the AV Public Transportation System \cite{AVPTS}, or to provide a kind of parking service to its subscribed AVs.  

Similar to many existing V2G implementations (e.g., \cite{V2G_capacity}), the system is considered to  operate in a time-slot basis. The time horizon is described  by time slots  $\{t=0,1,\ldots, D\}$.
As providing auxiliary services is one of the core functions of V2G  in which the extent of participation needs to be committed in advance in the corresponding auxiliary service markets, each parking facility $f$ is supposed to be able to estimate its demand profile $p_f=[\rho_t^f]_{1\leq t\leq D}$ by $t=0$. Moreover, with the advancement of vehicular communication technologies (e.g., vehicular ad-hoc networks \cite{VANET}),  the governed AVs are all connected and they can predict their travel plans for the near future. Thus it is possible for the system to determine the set of AVs with parking intention during the period  $\mathcal{T}=\{t=1,\ldots,D\}$  by $t=0$.\footnote{As AVs are more predictable, we assume that the availabilities of all AVs are known in advance. This is valid when it comes to dedicated transportation systems, e.g., the AV Public Transportation System \cite{AVPTS}. Moreover, we may adjust $D$ based on the amount of information about the AVs and parking facilities.} 
Therefore, we assume that all necessary information, from both the AVs and parking facilities, is available at $t=0$ and we will assign the AVs of $\mathcal{K}$ to appropriate parking facilities of $\mathcal{F}$ for the period $\mathcal{T}$.

\begin{figure}[!t]
\centering
\includegraphics[width=2.5in]{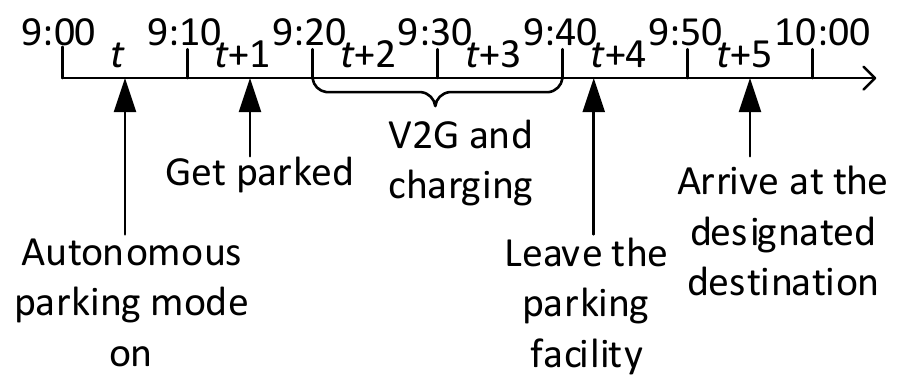} 
\caption{A time slot conversion example.}
\label{fig:timexample}
\end{figure}
To fit into the time-slot implementation of V2G, without loss of generality, we convert all the time parameters introduced in Sections \ref{subsec:AV} and \ref{subsec:PF}, including $\underline{t}_k, \overline{t}_k, \underline{m}_k,\overline{m}_k$, and $\hat{m}_k^f$,  to the time-slotted format.  Consider the scenario given in Fig. \ref{fig:timexample} which depicts the schedule of a particular AV $k$. \textcolor{black}{We set the duration of each time slot to  10 minutes for illustrative purposes}\footnote{\textcolor{black}{We will investigate the system performance with different time scales in Section \ref{sec:performance}.}} and the time slots start at 9:00, 9:10, 9:20, and so on. The AV is ready to park in time slot $t$ and it arrives at a parking facility in $t+1$. It leaves the parking facility in $t+4$ and returns to its designated destination in $t+5$. $k$ is only available for V2G and charging in $t+2$ and $t+3$. We can simply set $\underline{t}_k=t+1$, $\overline{t}_k=t+5$, and $\hat{m}_k^f=2$. It takes one slot for the first leg and one slot for the second leg of its journey, i.e., $\underline{m}_k=\overline{m}_k=1$. In this way, we have not only reserved sufficient time for $k$ to travel, but $k$ can also be made to fit into the V2G slotted operation.

After an assignment for the time horizon $\{t=0,1,\ldots,D\}$,  another assignment can be performed after time $\Delta t>0$, i.e., for $\{t=0+\Delta t, 1+\Delta t,\ldots, D+\Delta t\}$ . If $\Delta t$ is larger than $D$, it is like a fresh restart such that the two assignments have no correlation. If $\Delta t$ is smaller than $D$, it is possible that some AVs are still undergoing the schedules settled in the first assignment. We can still consider these AVs in the later assignment such that their parameters are revised to reflect their  updated statuses accordingly. For example,  if AV $k$ is parking at the parking facility $f$ at $t=\Delta t$, we may simply set its starting location to $\hat{n}_f$, i.e., $\underline{n}_f=\hat{n}_f$, for the later assignment.

\section{Problem Formulation} \label{sec:formulation}
To facilitate the formulation of the problem, we define two binary \textcolor{black}{variables} $x_{kt}^f$ and $y_k^f$ as follows:
\begin{align*}
x_{kt}^f= \begin{cases}
1 &\text{if AV $k$ is assigned to Parking Facility $f$} \\
    &\text{in the time slot $t$},\\
0 &\text{otherwise,}
\end{cases}
\end{align*}
and 
\begin{align*}
y_{k}^f= \begin{cases}
1 &\text{if AV $k$ is parked at $f$},\\
0 &\text{otherwise.}
\end{cases}
\end{align*}
Although $x_{kt}^f$ implies $y_k^f$, the introduction of $y_k^f$ can make the formulation simpler.

There are a number of requirements  governing the assignment of the AVs to the parking facilities. First, each AV should be allocated to a parking facility for proper parking. In general, an AV $k$ should not impose unnecessary burden to the traffic and should stay stationary in a parking facility most of the time from $\underline{t}_k$ to $\overline{t}_k$ . Hence, we consider that an AV will be assigned to one and only one parking facility during its off-duty period. This can be specified by  
\begin{align}
 \sum_{f\in\mathcal{F}}y_{k}^f=1,\forall k\in\mathcal{K}. 
 \label{con1}
\end{align}

If AV $k$ is assigned to Facility $f$, it will stay at $f$ for a sufficient number of time slots for charging and supporting V2G services. Recall that the parked duration $\hat{m}_k^f$  depends on its SOC specifications, the travel distances between its specific locations and $f$, and the expected utilization of $k$ for V2G by $f$.  When the details of $k$ and $f$ are given, by computing \eqref{AV2PF1} and \eqref{AV2PF2},  $\hat{m}_k^f$ is indeed a constant. We can represent such condition with the following inequality:
\begin{align}
	\hat{m}_k^{\textcolor{black}{f}} y_k^f\leq\sum_{t=1}^{D}x_{kt}^f \leq My_k^f, \forall k\in\mathcal{K},f\in\mathcal{F},  \label{eq:suffstay}
\end{align}
where $M$ is a sufficiently large positive number.

It takes time for an AV $k$ to travel from its original position $\underline{n}_k$ to a parking facility $f$ and return to  a designated location $\overline{n}_k$ after parking. The time periods for these two legs of journey are specified by $\underline{m}_k$ and $\overline{m}_k$, respectively (see Eq. \eqref{AV2PF1}). If  $k$ is parked at $f$ at time $t$, we should reserve at least  $\underline{m}_k$ time slots for $k$ to reach $f$. In other words, if $x_{kt}^f=1$, then there are at least $\underline{m}_k$ time slots with $x_{ks}^f=0$, where $s<t$. That is   $\sum_{s=\underline{t}_k}^{t-1}(1-x_{ks}^f)\geq \underline{m}_k$. This can be satisfied by imposing the following inequality:
\begin{align}
	  \sum_{s=\underline{t}_k}^{t-1}(1-x_{ks}^f)\geq \underline{m}_k x_{kt}^f, \forall k\in\mathcal{K}, f\in\mathcal{F},t\in\mathcal{T}.  \label{eq:totime}
\end{align}
Similarly, if $k$ is parked at $f$ at time $t$, we should reserve at least $\overline{m}_k$ time slots for $k$ to get back to $\overline{n}_k$ from $f$ by $\overline{t}_{k}$ for $x_{kt}^f=1$. This is equivalent to:
\begin{align}
	 \sum_{s=t+1}^{\overline{t}_k}(1-x_{ks}^f)\geq \overline{m}_k x_{kt}^f,  \nonumber\\
    \forall k\in\{k|\overline{t}_k-\overline{m}_k\leq D\}, f\in\mathcal{F},t\in\mathcal{T}. \label{eq:backtime}
\end{align}

An AV $k$ should be assigned to a facility $f$ such that its total travel distance does not exceed  $d_k^{max}$. In other words, if $y_k^f=1$, then $d_{\underline{n}_kf}+d_{f\overline{n}_k}\leq d_k^{max}$. This can be further described by:
\begin{align}
(d_{\underline{n}_k\hat{n}_f}+d_{\hat{n}_f\overline{n}_k}) y_k^f\leq d_k^{max} , \forall f\in\mathcal{F}, k\in\mathcal{K}. \label{eq:totaldistance} 
\end{align}

Since AV $k$ is available for parking from  $\underline{t}_k$ to $\overline{t}_k$ only, it should not be assigned to any parking facility any time before $\underline{t}_k$ and from $\overline{t}_k$ onward. This can be specified with the following two equalities:  
\begin{align}
	 \sum_{t=1}^{\underline{t}_k-1}x_{kt}^f=0,\forall f\in\mathcal{F},k\in\mathcal{K}  \label{eq:dutytime1}
\end{align}
and
\begin{align}
   \sum_{t=\overline{t}_k}^{D}x_{kt}^f=0,\forall f\in \mathcal{F},k\in\{k|\overline{t}_k-\overline{m}_k\leq D\}. \label{eq:dutytime2}
\end{align}
In fact, since $\underline{m}_k$ is known, we can combine \eqref{eq:totime} and \eqref{eq:dutytime1} resulting in
\begin{align}
 \sum_{t=1}^{\underline{t}_k-1+\underline{m}_k}x_{kt}^f=0,\forall f\in\mathcal{F},k\in\mathcal{K}. \label{eq:combinedto}
\end{align}
Similarly, combining \eqref{eq:backtime} and \eqref{eq:dutytime2} can get 
\begin{align}
   \sum_{t=\overline{t}_k-\overline{m}_k}^{D}x_{kt}^f=0,\forall f\in \mathcal{F},k\in\{k|\overline{t}_k-\overline{m}_k\leq D\}. 
\label{eq:combinedfrom}
\end{align}

To meet the demand from the V2G services, we should secure  enough AVs parked at $f$ based on its demand profile $p_f$. It is not uncommon to summarize the grid requirements with a total amount of energy required at each aggregator, e.g., in \cite{V2G_ancillary2}. We can also represent this amount of energy with a number of vehicles, each of which contributes equal portion, e.g., in \cite{V2G_capacity}. Moreover, the number of  AVs parked at $f$ should not exceed its capacity $c_f$. These can be ensured with the following inequality:
\begin{align}
 \rho_t^f\leq \sum_{k\in\mathcal{K}} x_{kt}^f \leq c_f , \forall f\in\mathcal{F},t\in\mathcal{T}. 
 \label{con6}
\end{align}
AVs should be parked as long as possible. We can do this by maximizing the occupancy, i.e., assigning the AVs to the parking facilities in as many time slots as possible. This is equivalent to maximizing $\sum_{k\in\mathcal{K},t\in\mathcal{T},f\in\mathcal{F}} x_{kt}^f $.\footnote{If economic cost of energy needs to be explicitly considered, we can simply replace the objective function with the related cost function.} We call the problem the Coordinated Parking Problem (CPP) and its formulation is given as follows:
\begin{problem}[Coordinated Parking Problem]
\label{prob:problem1}
\begin{equation}
\begin{aligned}
& \text{maximize} 
& & \sum_{k\in\mathcal{K},t\in\mathcal{T},f\in\mathcal{F}} x_{kt}^f \\
& \text{subject to} 
&& \eqref{con1}, \eqref{eq:suffstay}, \eqref{eq:totaldistance}, \eqref{eq:combinedto}\text{--}\eqref{con6}.
\end{aligned}
\label{problem1}
\end{equation}
\end{problem}
This problem is equivalent to the one formulated in \cite{smartgridcomm2016} but with much fewer constraints. This allows the problem to be solved more effectively. 
CPP is an integer linear program (ILP) and it can be solved by a standard ILP solver.

\section{Distributed Algorithm} \label{sec:distributed}
As will be shown in Section \ref{sec:performance}, if we solve CPP in a centralized manner, the computational time required grows tremendously with the number of AVs. In order to make it scalable, we are going to develop a distributed algorithm to speed up the computational process. We adopt the dual decomposition method \cite{bertsekas}, which have been widely applied to problems in power systems (e.g., \cite{cdc,voltage}), to develop the distributed algorithm. 

Based on Problem \ref{prob:problem1}, we first relax Constraint \eqref{con6} by introducing \textcolor{black}{Lagrangian} multipliers $\overline{\lambda}_t^f$ and $\underline{\lambda}_t^f$ and construct the partial Lagrangian as follows:
\begin{align*}
& \sum_{k\in\mathcal{K},t\in\mathcal{T},f\in\mathcal{F}} x_{kt}^f - \sum_{t\in\mathcal{T},f\in\mathcal{F}}\overline{\lambda}_t^f\left(\sum_{k\in\mathcal{K}}x_{kt}^f - c_f\right)\\
& - \sum_{t\in\mathcal{T},f\in\mathcal{F}}\underline{\lambda}_t^f\left(-\sum_{k\in\mathcal{K}}x_{kt}^f + \rho_t^f\right)\\
=& \sum_{k\in\mathcal{K},t\in\mathcal{T},f\in\mathcal{F}} x_{kt}^f -
\sum_{k\in\mathcal{K},t\in\mathcal{T},f\in\mathcal{F}}\overline{\lambda}_t^fx_{kt}^f\\
&+\sum_{k\in\mathcal{K},t\in\mathcal{T},f\in\mathcal{F}}\underline{\lambda}_t^fx_{kt}^f+
\sum_{t\in\mathcal{T},f\in\mathcal{F}}\overline{\lambda}_t^fc_f - 
\sum_{t\in\mathcal{T},f\in\mathcal{F}}\underline{\lambda}_t^f\rho_t^f\\
=& \sum_{k\in\mathcal{K},t\in\mathcal{T},f\in\mathcal{F}}(x_{kt}^f-\overline{\lambda}_t^fx_{kt}^f+\underline{\lambda}_t^fx_{kt}^f) \\
&+ \sum_{t\in\mathcal{T},f\in\mathcal{F}} (\overline{\lambda}_t^fc_f-\underline{\lambda}_t^f\rho_t^f).
\end{align*}

Clearly the rest of the constraints, i.e., \eqref{con1}, \eqref{eq:suffstay}, \eqref{eq:totaldistance}, \eqref{eq:combinedto}\text{--}\eqref{eq:combinedfrom}, are all separable with respect to $k$. For each $k$, we represent the variables and feasible region confined by \eqref{con1}, \eqref{eq:suffstay}, \eqref{eq:totaldistance}, \eqref{eq:combinedto}\text{--}\eqref{eq:combinedfrom} as $\sigma_k = \{x_{kt}^f,y_k^f\}_{t\in\mathcal{T},f\in\mathcal{F}}$ and $\mathcal{Z}_k$, respectively. Let $\overline{\Lambda}=\{\overline{\lambda}_t^f\}_{t\in\mathcal{T},f\in\mathcal{F}}$ and $\underline{\Lambda}=\{\underline{\lambda}_t^f\}_{t\in\mathcal{T},f\in\mathcal{F}}$. Thus the dual function $g(\overline{\Lambda},\underline{\Lambda})$ of Problem \ref{prob:problem1} becomes
\begin{align}
g(\overline{\Lambda},\underline{\Lambda})=&\sum_{k\in\mathcal{K}}\sup_{\sigma_k\in\mathcal{Z}_k}\left\{\sum_{t\in\mathcal{T},f\in\mathcal{F}}(x_{kt}^f-\overline{\lambda}_t^fx_{kt}^f+\underline{\lambda}_t^fx_{kt}^f)\right\} \nonumber\\
&+ \sum_{t\in\mathcal{T},f\in\mathcal{F}} (\overline{\lambda}_t^fc_f-\underline{\lambda}_t^f\rho_t^f),
\label{eq:dualfunction}
\end{align}
which is convex because of the pointwise supremum of affine functions of $(\overline{\Lambda},\underline{\Lambda})$.
We can also see that the first summation of \eqref{eq:dualfunction} clearly decouples with respect to $k$.
Given $(\overline{\Lambda},\underline{\Lambda})$, we define the subproblem for each $k\in\mathcal{K}$ as follows:
\begin{problem}[Subproblem for AV $k$]
\label{subprob}
\begin{subequations}
\label{subproblem1}
\begin{align}
 \text{maximize} 
 & \sum_{t\in\mathcal{T},f\in\mathcal{F}}(x_{kt}^f-\overline{\lambda}_t^fx_{kt}^f+\underline{\lambda}_t^fx_{kt}^f)\\
 \text{subject to} 
& \sum_{f\in\mathcal{F}}y_{k}^f = 1\\
& \hat{m}_k^{\textcolor{black}{f}} y_k^f\leq\sum_{t=1}^{D}x_{kt}^f \leq My_k^f, \forall f\in\mathcal{F},\\
& (d_{\underline{n}_k\hat{n}_f}+d_{\hat{n}_f\overline{n}_k}) y_k^f\leq d_k^{max} , \forall f\in\mathcal{F},\\
& \sum_{t=1}^{\underline{t}_k-1+\underline{m}_k}x_{kt}^f=0,\forall f\in\mathcal{F},\\
& \sum_{t=\overline{t}_k-\overline{m}_k}^{D}x_{kt}^f=0,\forall f\in \mathcal{F}. \label{eq:18f}
\end{align}
\end{subequations}
\end{problem}
\hspace{-0.3cm}For those $k$ with $\overline{t}_k-\overline{m}_k> D$, \eqref{eq:18f} can be ignored.
Let $g_k(\overline{\Lambda},\underline{\Lambda})$ be the optimal value of \eqref{subproblem1} for $k$. We update the dual variables $\overline{\Lambda}$ and $\underline{\Lambda}$ by addressing the dual problem:
\begin{subequations}
\label{subproblem}
\begin{align}
& \text{minimize} \quad
 & \sum_{k\in\mathcal{K}}g_k(\overline{\Lambda},\underline{\Lambda})+\sum_{t\in\mathcal{T},f\in\mathcal{F}} (\overline{\lambda}_t^fc_f-\underline{\lambda}_t^f\rho_t^f)\\
& \text{subject to} 
& \overline{\Lambda},\underline{\Lambda}\geq 0,
\end{align}
\end{subequations}
which is linear. We can solve the dual problem to recover the solution of the original Problem \ref{prob:problem1}. 
We have the gradients $\frac{\partial g_k(\overline{\Lambda},\underline{\Lambda})}{\partial \overline{\lambda}_t^f} = c_f - \sum_{k\in\mathcal{K}}x_{kt}^f(k)$
and
$\frac{\partial g_k(\overline{\Lambda},\underline{\Lambda})}{\partial \underline{\lambda}_t^f} =  \sum_{k\in\mathcal{K}}x_{kt}^f(k) - \rho_t^f$. By projected gradient descent \cite{cvx}, we can generate a sequence of feasible points $\{\overline{\Lambda}(i),\underline{\Lambda}(i)\}$ with the following update rules:
\begin{align}
\overline{\lambda}_{t}^f(i+1) = \left[\overline{\lambda}_{t}^f(i)-\overline{\gamma}_t^f(i)\left(  c_f-\sum_{k\in\mathcal{K}}x_{kt}^{f*}(i)\right)\right]^+, \nonumber\\
\forall t\in\mathcal{T},f\in\mathcal{F}, \label{update1}\\
\underline{\lambda}_{t}^f(i+1) = \left[\underline{\lambda}_{t}^f(i)-\underline{\gamma}_t^f(i)\left(\sum_{k\in\mathcal{K}}x_{kt}^{f*}(i) - \rho^f_t\right)\right]^+, \nonumber\\
\forall t\in\mathcal{T},f\in\mathcal{F},\label{update2}
\end{align}
where $x_{kt}^{f*}(i)$ is the optimal result by solving Problem \ref{subprob} at Iteration $i$ while $\overline{\gamma}_t^f(i)>0$ and $\underline{\gamma}_t^f(i)>0$ are the step sizes at Iteration $i$. If we have $\sum_{k\in\mathcal{K}}x_{kt}^{f*}(i)> c_f$ violating \eqref{con6}, \eqref{update1} will make $\overline{\lambda}_{t}^f(i+1)>\overline{\lambda}_{t}^f(i)$. Solving Problem \ref{subprob} at Iteration $i+1$ tends to make $x_{kt}^{f*}(i+1)$ smaller. Similarly, if we have $\sum_{k\in\mathcal{K}}x_{kt}^{f*}(i)< \rho_t^f$ violating \eqref{con6},  \eqref{update2} will make $\underline{\lambda}_{t}^f(i+1)>\underline{\lambda}_{t}^f(i)$. Solving Problem \ref{subprob} at Iteration $i+1$ tends to make $x_{kt}^{f*}(i+1)$ larger.
We can interpret $(\overline{\Lambda},\underline{\Lambda})$ as a set of shadow prices for the parking resources: $\overline{\lambda}_t^f$ and $\underline{\lambda}_t^f$ are the price of renting a parking space and the price of selling V2G services at parking facility $f$ in time slot $t$, respectively.\footnote{\textcolor{black}{Note that the shadow prices serve to provide another way to interpret \eqref{update1} and \eqref{update2} economically only.}} On one hand, if the number of required parking slots is larger than the capacity (i.e., $\sum_{k\in\mathcal{K}}x_{kt}^{f*}> c_f$), the parking space selling price (i.e., $\overline{\lambda}_{t}^f$) will increase and this may lower the total demand $\sum_{k\in\mathcal{K}}x_{kt}^{f*}(i)$. On the other hand, if the number of AVs contributing to V2G is smaller than the energy profile (i.e., $\sum_{k\in\mathcal{K}}x_{kt}^{f*}< \rho_t^f$), the V2G service charge $\underline{\lambda}_t^f$ will increase and this encourages more AVs to park at $f$ in time slot $t$. As a whole, the dual problem is used to control the shadow prices and each AV adjusts its own parking strategy with the subproblem based on the parking fees $\overline{\Lambda}$ and the V2G service charges $\underline{\Lambda}$.

\begin{figure}
\centering
\includegraphics[width=2.4in]{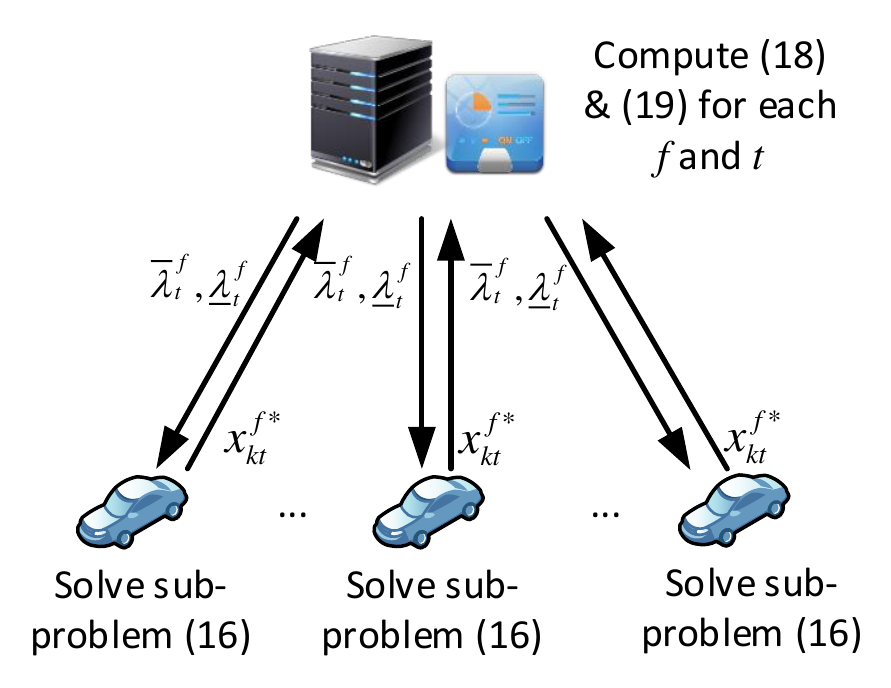}
\caption{Implementation of the distributed algorithm.}
\label{fig:distributed}
\end{figure}

Suppose there is a control center which manages the whole system. Fig. \ref{fig:distributed} depicts how to implement the distributed algorithm. In terms of computation, the control center updates $(\overline{\Lambda},\underline{\Lambda})$ with \eqref{update1} and \eqref{update2} while each AV solves its own suproblem, i.e., Problem \ref{subprob}. After updating $(\overline{\Lambda},\underline{\Lambda})$, the control center distributes $(\overline{\Lambda},\underline{\Lambda})$ to the AVs. Similarly, after solving the subproblem, each AV passes its optimal $x_{kt}^{f*}$ to the control center. The overall distributed algorithm is illustrated by Algorithm \ref{alg:DA}. 
\begin{algorithm}
\small
\caption{Distributed Algorithm}
\label{alg:DA}
\begin{tabbing}
1. \=Initialize $\overline{\Lambda}$ and $\underline{\Lambda}$\\
2. \textbf{while} stopping criteria not matched \textbf{do}\\
\>3. \textbf{for} \= each AV $k$ (in parallel) \textbf{do}\\
\>\>4. Given $\overline{\Lambda}$ and $\underline{\Lambda}$, solve \eqref{subproblem1}\\
\>\>5. Return $x_{kt}^{f*}, \forall t\in\mathcal{T},f\in\mathcal{F}$\\
\>6. \textbf{end for}\\
\>7. \textbf{for} \= each $t\in\mathcal{T},f\in\mathcal{F}$ (in parallel) \textbf{do}\\
\>\>8. Given $x_{kt}^{f*}, \forall t\in\mathcal{T},f\in\mathcal{F}$, update each $\overline{\lambda}_t^f$ and $\underline{\lambda}_t^f$ with\\ 
\>\>\quad \eqref{update1} and \eqref{update2}, respectively\\
\>\>9. Distribute $\overline{\Lambda}$ and $\underline{\Lambda}$ to the AVs\\
\>10. \textbf{end for}\\
11. \textbf{end while}
\end{tabbing}
\end{algorithm}

We first initialize $(\overline{\Lambda},\underline{\Lambda})$ with appropriate non-negative values at the control center (Step 1). Then the algorithm iterates until a stopping criterion has been satisfied (Steps 2--11). Each iteration is divided into two parts. The first part (Steps 3--6) corresponds to solving the subproblems. After receiving $(\overline{\Lambda},\underline{\Lambda})$, each AV solves \eqref{subproblem1} in parallel and returns the computed $x_{kt}^{f*}$'s to the control center (Step 5). The second part (Steps 7-10) is for updating $(\overline{\Lambda},\underline{\Lambda})$. After collecting the $x_{kt}^{f*}$'s for particular $t$ and $f$, the control center can compute the corresponding $\overline{\lambda}_t^f$ and $\underline{\lambda}_t^f$ (Step 8). Hence updating $(\overline{\Lambda},\underline{\Lambda})$ can be done in parallel. The resultant $(\overline{\Lambda},\underline{\Lambda})$ is then distributed to the AVs (Step 9). Suppose $g_k(i)$ is the optimal value of the subproblem for AV $k$ at iteration $i$. We consider the algorithm converged if 
\begin{align}
\frac{|\sum_kg_k(i+1)-\sum_kg_k(i)|}{|\sum_kg_k(i+1)|}<\delta, \forall k\in\mathcal{K},
\label{eq:stopping}
\end{align}
where $\delta$ is a small positive value, e.g., $10^{-5}$.

The primal solution of the original CPP can be retained by the solutions of the subproblems collectively.
We can recover the primal solution from the dual as follows:
We first determine the $(t,f)$ pair which has the largest AV deficit, i.e., $\rho^f_t-\sum_{k\in\mathcal{K}}x^f_{kt}$. Then we construct a list of ``free'' AVs which can be moved to $f$ at $t$. Each AV $k^\prime$ in the list can be removed from its original parking facility $f^\prime$ without violating the respective energy profile constraint $\rho^{f^\prime}_t-\sum_{k\in\mathcal{K}\setminus k^\prime}x^{f^\prime}_{kt}\leq0,\forall t\in\{t|x^{f^\prime}_{k^\prime t}=1\}$,
and the AV must be able to park in $f$ at $t$, i.e., $\underline{t}_k+\underline{m}_k\leq t < \overline{t}_k-\overline{m}_k$.
Among those AVs in the list, the one with longest possible stay in $f$,  calculated by $\overline{t}_k-\overline{m}_k-(\underline{t}_k+\underline{m}_k)$, is selected to park in $f$ from $\underline{t}_k+\underline{m}_k$ to $\overline{t}_k-\overline{m}_k-1$. Thus $x^{f^\prime}_{k^* t} = 0,\forall t\in\mathcal{T}$ and $x^{f}_{k^* t} = 0,\forall t\in[\underline{t}_{k^*}+\underline{m}_{k^*}, \overline{t}_{k^*}-\overline{m}_{k^*})$, where $k^*$ is the selected AV. 
The parking capacity constraint is handled in a similar manner. The $(t,f)$ pair with the largest AV overflow is identified by calculating $\sum_{k\in\mathcal{K}}x^f_{kt} - c_f$. A list of ``free'' AVs that can be removed without violating the respective energy profile constraint is developed. Then the AV with shortest possible stay is removed from $f$. A feasible primal solution is generated when $\rho^f_t-\sum_{k\in\mathcal{K}}x^f_{kt}\leq0,\forall t\in\mathcal{T}, k\in\mathcal{K}$ and $c_f-\sum_{k\in\mathcal{K}}x^f_{kt}\geq0,$ for all $ t\in\mathcal{T}, k\in\mathcal{K}$.

\textcolor{black}{
Algorithm \ref{alg:DA} requires the minimum amount of information exchange. In each iteration, after receiving the pricing signals from the control center, each AV addresses its own subproblem with AV-specific parameters (including $\hat{m}_k^f$, $d_{\underline{n}_k\hat{n}_f}$, $d_{\hat{n}_f\overline{n}_k}$, and $d_k^{max}$) only. After receving the AVs' preferences on the parking facility assignments (in terms of $x_{kt}^{f*}$), the control center updates the shadow prices with the parking facility-specific parameters ($c_f$ and $\rho_t^f$) only. In a practical system, the number of AVs should be far more than the number of parking facilities. Asking each vehicle to handle its own subproblem with their own parameters avoids gathering many scattered vehicular data, which make the method highly practical. 
}

\section{Performance Evaluation} \label{sec:performance}

We have developed three methods to solve CPP, namely (I) centralized, (II) heuristic, and (III) distributed approaches. With Method I, we directly apply a standard ILP solver to Problem \ref{prob:problem1} and we adopt Gurobi \cite{gurobi-ref} here. Method II is illustrated in \cite{smartgridcomm2016} while Method III has been introduced in Section \ref{sec:distributed}.

We perform four tests to evaluate the performance of the solution methods, with emphasis on Method III. In the first test, we assess their performance on different scales of the problem with different numbers of AVs and parking facilities. The second test aims to investigate the effect of time scaling while the third test examines the convergence of Method III. In the fourth test, we study the performance of the distributed algorithm in the presence of communication loss.
We generate random cases \textcolor{black}{for testing}.
Unless stated otherwise, we assume that there are 100 time slots (i.e., $D=100$) evenly distributed in a horizon of two hours.
Consider a residential area of $5 \times 5$ $\text{km}^2$, 
within which we randomly place required numbers of AVs and parking facilities by specifying $\underline{n}_k$, $\overline{n}_k$, and $\hat{n}_f$ accordingly. Suppose that the AVs travel at a constant speed of 30 km/h. For AV $k$, the travel times spent on the two legs for parking, i.e., $\underline{m}_k$ and $\overline{m}_k$, are assigned based on the corresponding distances. 
We also set $\hat{m}_k^f=rand(1, \overline{t}_k-\overline{m}_k-(\underline{t}_k+\underline{m}_k))$. These capture $\alpha_k$ and $\beta_f$.
We specify $\underline{t}_k$ and $\overline{t}_k$ by $\underline{t}_k=\mathrm{rand}(0,D-\underline{m}_k-\overline{m}_k)$ and $\overline{t}_k=\mathrm{rand}(0,D-\underline{m}_k-\overline{m}_k)+\underline{t}_k+\underline{m}_k+\overline{m}_k$, where $\mathrm{rand}(\cdot, \cdot)$ produces 
an integer uniformly distributed between the two inputs inclusively. 
If $\overline{t}_k > D$, then AV $k$ will not need to return to  $\overline{n}_k$ during the time horizon. 
\textcolor{black}{$d_k^{max}$ is randomly set in the range of $[4,5]$ km.}
Finally, the energy profile of Parking Facility $f$ is set as $\rho_t^f=\mathrm{rand}(0,a_t^f/|\mathcal{F}|),\forall t\in \mathcal{T},$
where $a_t^f$ is the number of AVs that are available to park in $f$ at $t$.
The parking capacity $c_f$ is set to $|\mathcal{K}|/2$ for all $f$.
This allows us to generate feasible instances more easily to inspect the computational abilities of the methods.
All simulations are performed on a computer with Intel Core-i5 CPU at 2.90 GHz with 8 GB RAM. The simulations are coded with Python on Linux.

\subsection{Implementation of the Distributed Algorithm}
As Method III is implemented distributedly in each iteration (see Algorithm \ref{alg:DA}), the subproblem which takes the longest time contributes the time needed for the first part of an iteration while the update of the $\overline{\lambda}_t^f$ and $\underline{\lambda}_t^f$ which needs the longest time contributes the second part. As only small messages containing $x_{kt}^f$ or $(\overline{\Lambda},\underline{\Lambda})$ need to be passed among the entities, the communication delay \textcolor{black}{should be small. Based on the average latency of practical cellular systems \cite{latency}, we assume each iteration takes 200 ms of communication delay. }

We set $\delta$ in \eqref{eq:stopping} to $10^{-5}$. For all $f$ and $t$, we initialize $\overline{\lambda}_t^f$, $\underline{\lambda}_t^f$, $\overline{\gamma}_t^f$, $\underline{\gamma}_t^f$ with 0, 0, 0.01, and 0.01, respectively.
In a subsequent iteration $i$, we get $\overline{\gamma}_t^f(i+1)=\overline{\gamma}_t^f(i)\times 1.1$ if $\sum_kg_k(i)-\sum_kg_k(i-1)<0$. Otherwise, $\overline{\gamma}_t^f(i+1)=\overline{\gamma}_t^f(i)\times 0.1$. In addition, we introduce $\gamma^\text{cap}(i)$ to cap the step size such that $\overline{\gamma}_t^f(i)\leq\gamma^\text{cap}(i)$. We set  $\gamma^\text{cap}(i)=\gamma^\text{init}(1-\epsilon)^i$, where $\epsilon = 10^{-3}$ and $\gamma^\text{init} = 0.01$. This satisfies the nonsummable diminishing step size rule which guarantees the convergence of the algorithm \cite{bertsekas}. We modify $\underline{\gamma}_t^f$ similarly.

\subsection{Test 1: Different Scales of the Problem}




\begin{figure}
\centering
\includegraphics[width=3.2in]{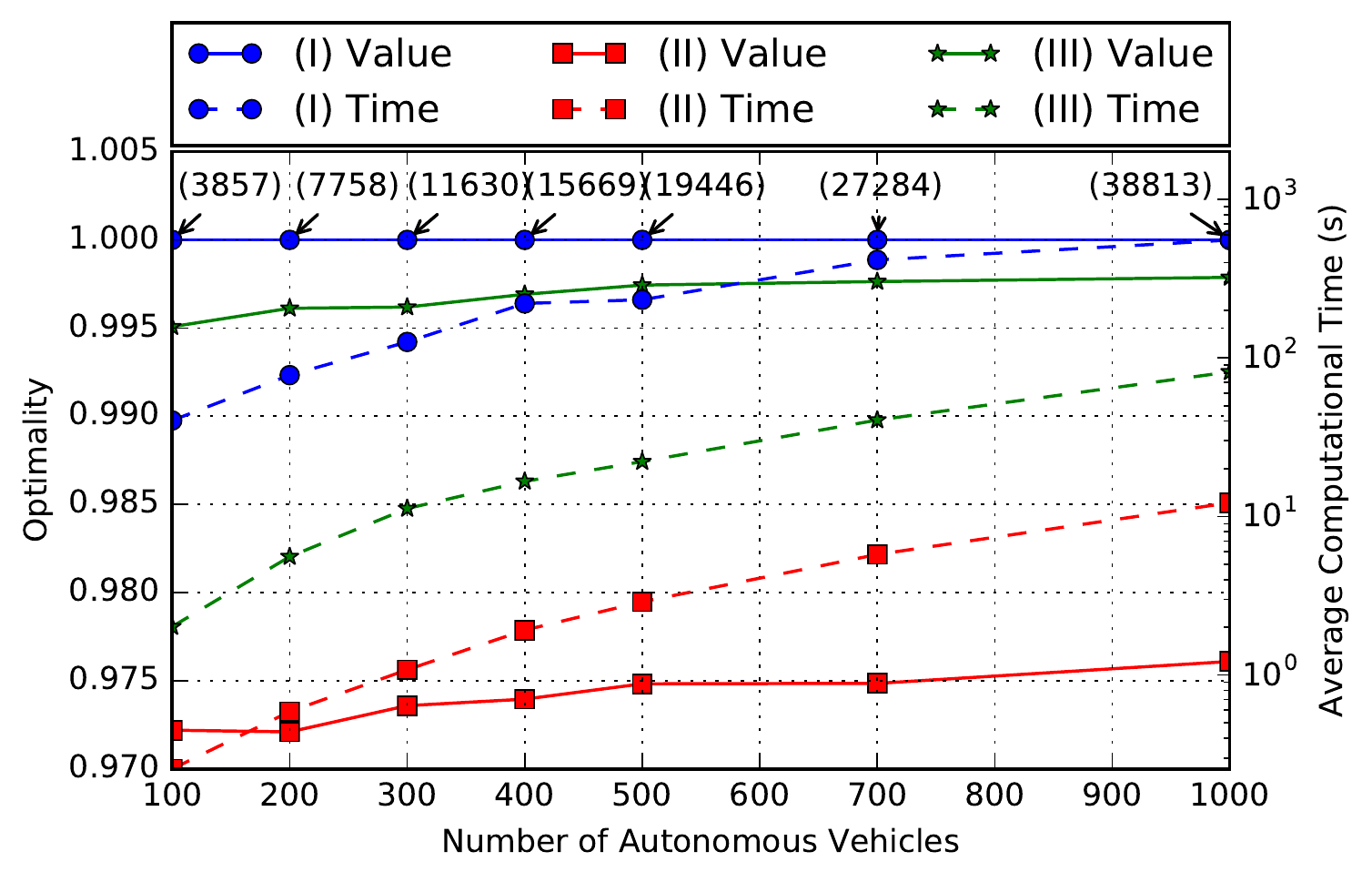} 
\caption{Objective function values and computation times with different numbers of AVs.}
\label{fig:compTimes}
\end{figure}
\begin{figure}
\centering
\includegraphics[width=3.2in]{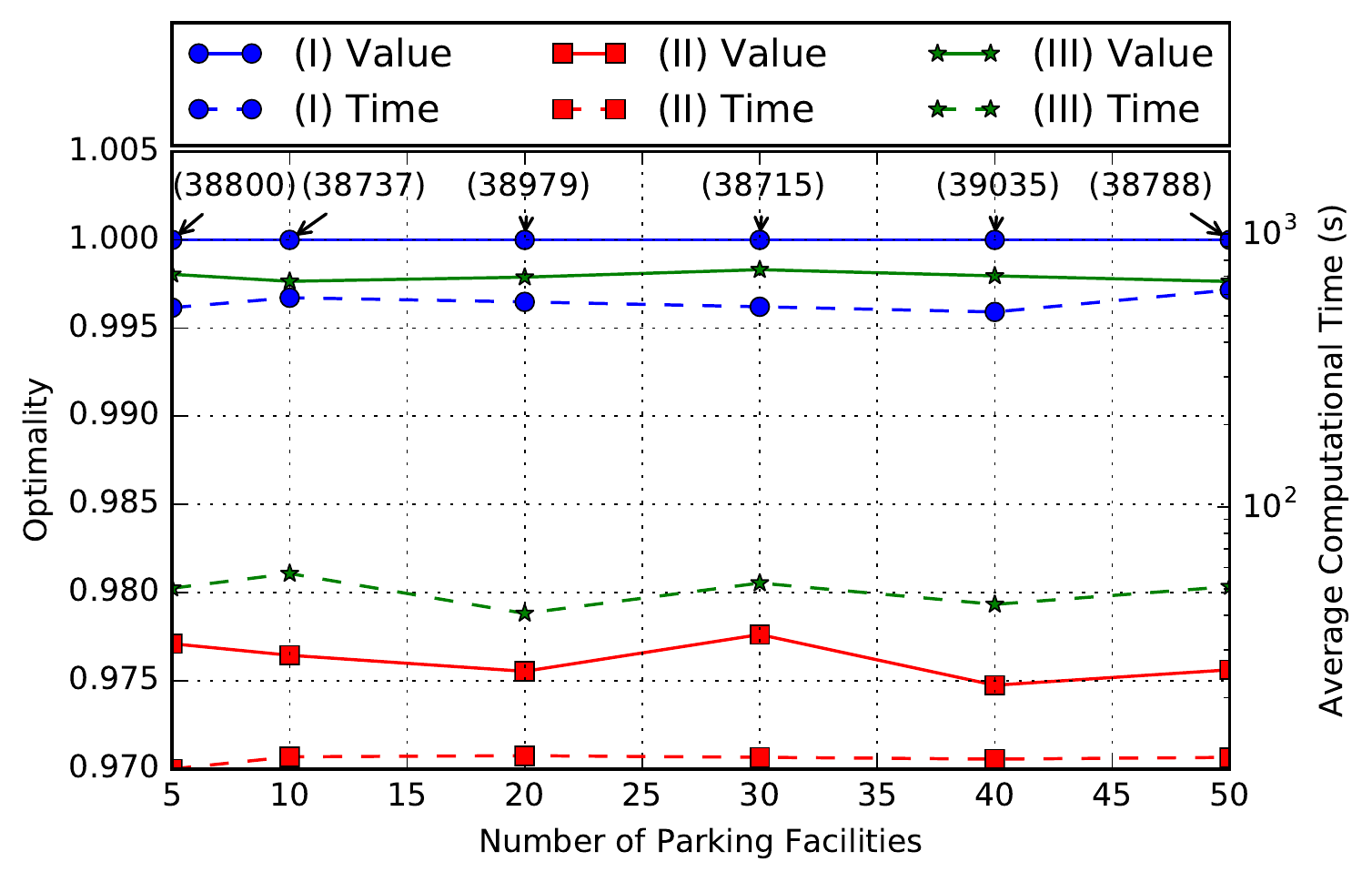} 
\caption{Objective function values and computation times with different numbers of parking facilities.}
\label{fig:compTimes2}
\end{figure}

We first examine different numbers of AVs with a fixed number of parking facilities. We consider a setting for a small neighborhood, where there are five parking facilities. We generate random cases of 100, 200, 300, 400, 500, 700, and 1000 AVs. Fig. \ref{fig:compTimes} depicts the relative objective function values (i.e., occupancies) with respect to the optimal and computational times (in log scale) obtained by the three methods. 
\textcolor{black}{The actual optimal values are also shown in brackets for reference.}
Each point in the figure corresponds to the average results from 25 cases. It can be observed that Method I always produces the optimal solutions while Method II can generate sub-optimal solutions which are about 97\% from the optimal. Method III is a little inferior to Method I but much better than Method II. Although both Methods II and III cannot guarantee optimality, they can produce better solutions in those cases with more AVs. The reason is that more AVs provide larger flexibility and it is easier for the algorithms to generate better solutions.
All methods need more computational time when the number of AVs grows. Method I is the most time demanding while Method III needs the shortest amount. Therefore, if the true optimal is needed, we will go for Method I, but its computational time grows significantly with problem size. Method III is very effective in producing high quality solutions and suitable for practical situations.

We further consider different numbers of parking facilities with a fixed number of AVs. We fix the number of AVs to 1000 and consider cases of 5, 10, 20, 30, 40, and 50 parking facilities. Similarly, Fig. \ref{fig:compTimes2} gives the relative computed objective function values with respect to the optimal and computational times, where each data point represents the average of 25 cases. 
\textcolor{black}{The actual optimal values are also shown in brackets for reference.}
We can see that both the objective function value and computational time are not very sensitive to the number of parking facilities. Thanks to the fact that the occupancy of a vehicle at a parking facility in a time slot has no difference from any other in the objective function, as long as the parking facilities are sufficient to accommodate the AVs, more parking facilities available will not help improve the objective function value. We can understand the trend of computational time in a similar way.

\subsection{Test 2: Time Scaling} \label{subsec:time}
Here we investigate the impact of time scaling. Recall that a given time horizon is divided into slots and we can make the division finer with more time slots for the same period. We generate 10 random cases for the same horizon of two hours. For each case, we divide the horizon into 10, 20, 30, 40, 50, 80, and 100 time slots with the same settings of 100 AVs and 5 parking facilities. In other words, we are solving the same problem instances with different time scales only. For example, the \textcolor{black}{$51^\text{st}$} time slot in the 100-scale corresponds to the $26^\text{th}$ and $5^\text{th}$ in the 50- and 10-scale, respectively. Since time scaling is intrinsic to the problem, we demonstrate its effects on the optimality and thus we show the results here with Method I only. Table I illustrates the percentage (\%) optimality of the different scales with respect to the 100-scale for the 10 cases. 
The 100-scale is the finest and gives the best results in term of quality.
When scaling down, the \% optimality drops slightly because the flexibility of assignment decreases. However, too coarse scaling (e.g., 10-scale) can result in infeasible solutions. Table I also shows the computational times averaged over the feasible cases. This suggests that scaling-down can improve the computational time significantly due to the reduced problem size. Therefore, there exists a tradeoff between solution quality and computational time.

\begin{table}
\renewcommand{\arraystretch}{1.2}
\centering
\label{tab:optimality}
\caption{Effects of time scaling on \% Optimality and computational time.}
\begin{tabular}{c|cccccccc}
\hline\hline
Case & 10 & 20 & 30 & 40 & 50 & 80 & 100\\\hline
I & N/A & 94.27 & 95.32 & 97.14 & 98.83 &  97.54  & 100.00\\
II & N/A & N/A & N/A & 96.97 & 98.67 &   98.10   & 100.00\\ 
III & N/A & 95.54 & 95.98 & 96.87 & 98.73 &  97.47  & 100.00\\
IV & N/A & N/A & 96.72 & 98.01 & 99.79 &   97.74  & 100.00\\
V & N/A & N/A & 96.04 & 97.26 & 99.00 &   97.85& 100.00\\
VI & N/A & N/A & 95.29 & 97.39 & 99.55 &   97.62 & 100.00\\
VII & N/A & N/A & 96.73 & 96.75 & 98.66 &   98.21  & 100.00\\
VIII & N/A & 94.72 & 95.28 & 96.19 & 98.93 &  97.32  & 100.00\\
IX & N/A & 94.49 & 95.62 & 96.65 & 98.46 &  97.82  & 100.00\\
X & N/A & 94.94 & 95.83 & 96.83 & 98.65 &  97.84 & 100.00\\ \hline
Avg. & \multirow{ 2}{*}{N/A} & \multirow{ 2}{*}{1.41}	& \multirow{ 2}{*}{2.08}	& \multirow{ 2}{*}{2.43}	& \multirow{ 2}{*}{3.17} & \multirow{ 2}{*}{5.30}	& \multirow{ 2}{*}{9.73}\\
time (s)\\
\hline\hline
\end{tabular}
\end{table}

\subsection{Test 3: Convergence of the Distributed Algorithm}
\begin{figure}
\centering
\includegraphics[width=3.2in]{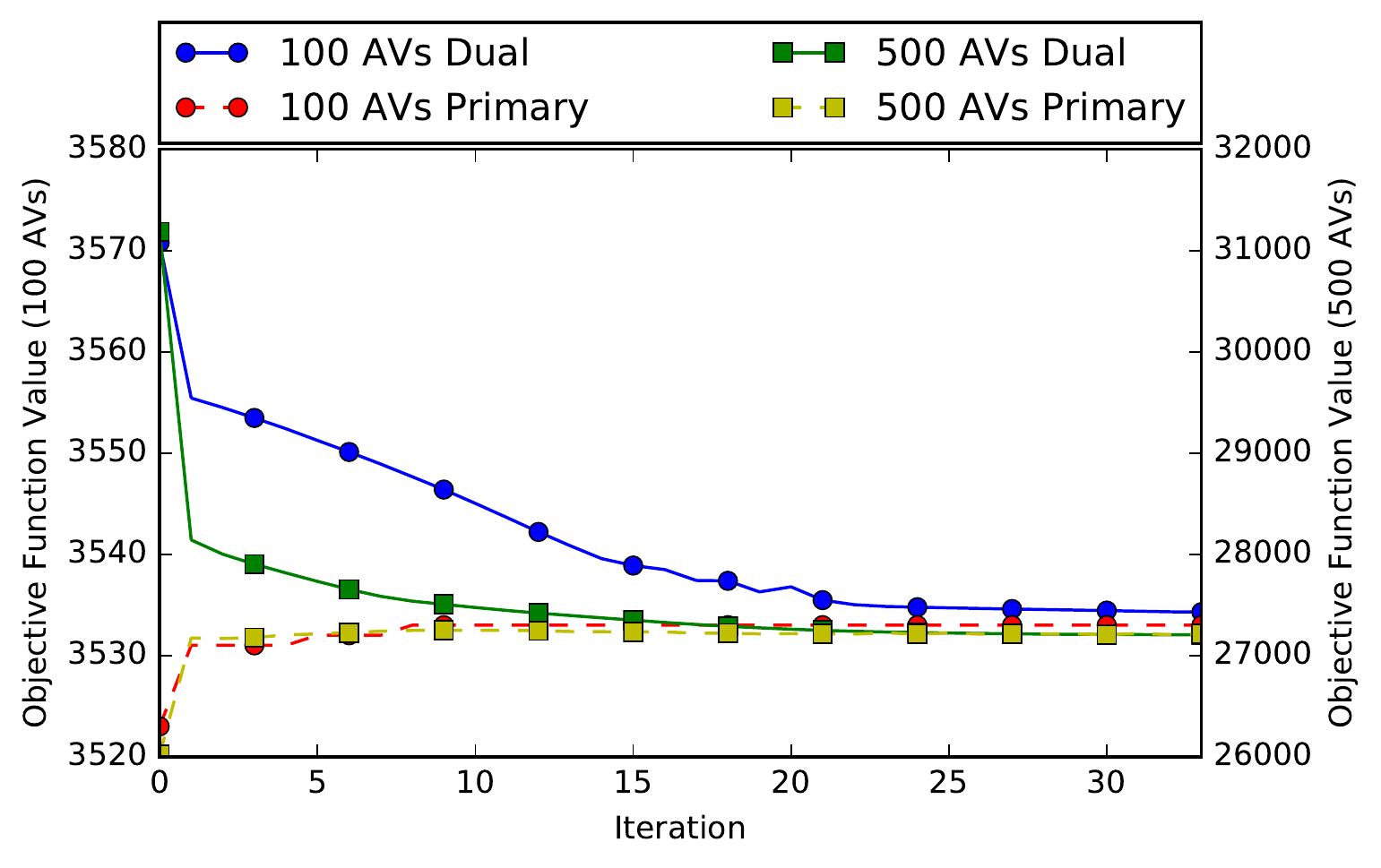} 
\caption{Primal and dual convergence of the distributed algorithm.}
\label{fig:convergence}
\end{figure}

In this test, we study the convergence of the distributed algorithm. For illustrative purposes, we examine two representative cases of 100 and 500 AVs with five parking facilities, respectively, from Test 1, both of which are accommodated by five parking facilities. Recall that the algorithm manipulates the dual solutions and we can recover the corresponding primal solutions with the method discussed in Section \ref{sec:distributed}. Fig. \ref{fig:convergence} shows the objective function values of the corresponding primal and dual solutions in different iterations. For both cases, the duality gaps diminish when the algorithm iterates. It converges faster in the larger case and this is consistent with the results given in Fig. \ref{fig:compTimes}. 

\subsection{Test 4: Communication Loss}
\begin{figure}
\centering
\includegraphics[width=3.0in]{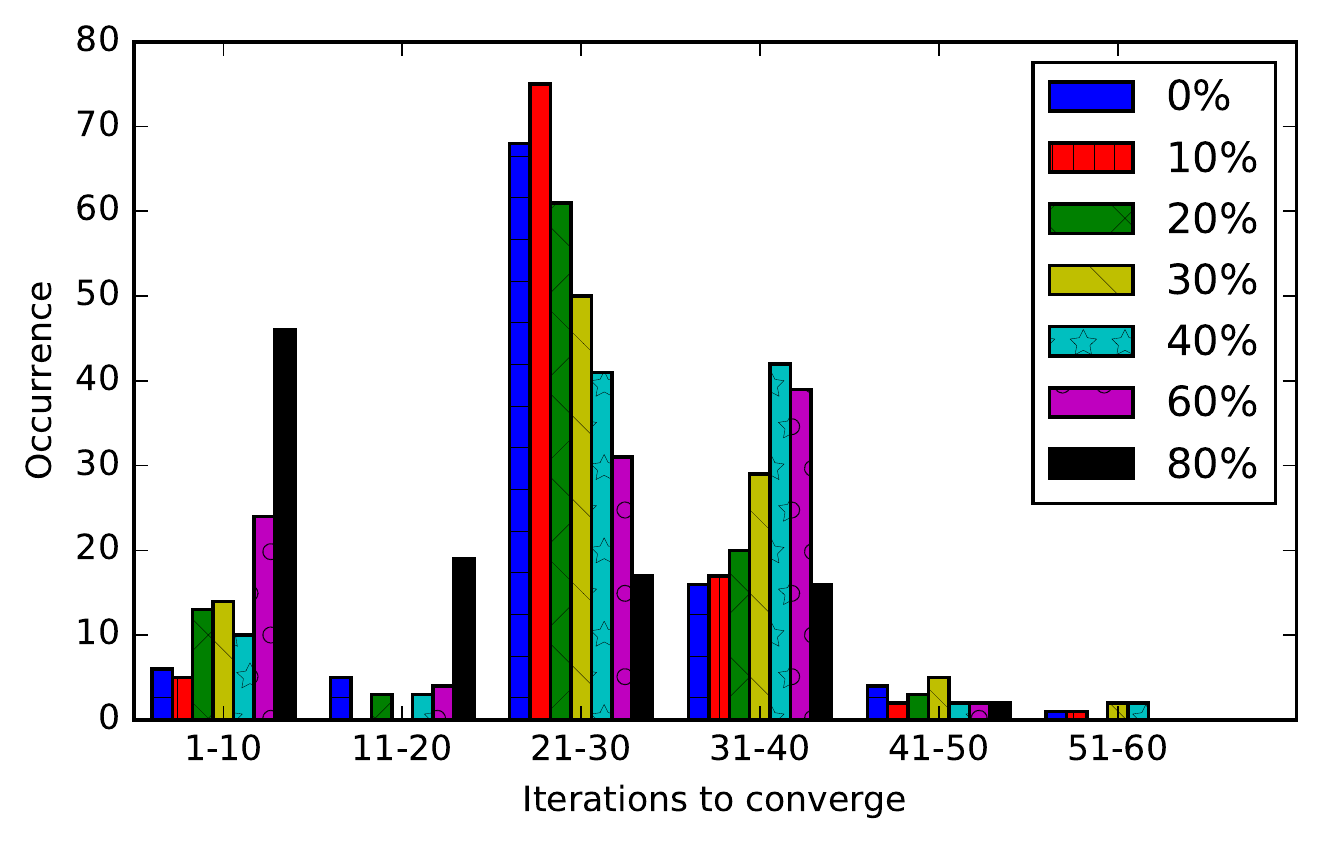} 
\caption{Iterations needed for convergence with different amount of communication loss.}
\label{fig:loss}
\end{figure}

Here we evaluate the performance of the distributed algorithm with the presence of communication loss. Recall that the algorithm relies on message passing to drive its convergence. Messages are passed around different entities in a communication network in the form of data packets. However, some packets may be lost during the transmission. We define $p$ as the probability of having a packet drop. When experiencing a packet drop, the involved entity uses the most recently received $x_{kt}^f$ or $(\overline{\Lambda},\underline{\Lambda})$ to do the calculation. We consider $p$ equal to 0\%, 10\%, 20\%, 30\%, 40\%, \textcolor{black}{60\%, and 80\%}, and for each of which we produce 100 random cases.
Fig. \ref{fig:loss} illustrates the occurrence of the 100 cases for each $p$ with respect to the number of iterations required for the algorithm to converge \textcolor{black}{and Table II indicates the corresponding maximum number of iterations among the 100 cases for each $p$.} While 40\% communication loss results in slightly slower convergence, severer communication loss does not make significant degradation in performance in general. 

\begin{table}\label{tab:lossss}
\renewcommand{\arraystretch}{1.2}
\centering
\caption{Maximum number of iterations required to converge.}
\textcolor{black}{
\begin{tabular}{c|cccccccc}
\hline\hline
\% Loss	& 0 & 10 & 20 & 30 & 40 & 60 & 80 \\\hline
No. of iterations & 50 & 54 & 46 & 56 &  55  & 45 & 49\\
\hline\hline
\end{tabular}
}
\end{table}


\section{Conclusion} \label{sec:conclusion}
AVs will represent a substantial share of ground transport in the near future. When parked, AVs can participate in V2G as EVs do. The difference is that AVs can be instructed to travel based on some system-wide objectives. In this paper, we study how to coordinate AVs intending to park, to reach parking facilities for supporting V2G services. We formulate CPP in the form of ILP. Besides  solving it in a centralized manner by a standard ILP solver, we propose a distributed algorithm to overcome efficiency issue of the centralized approach. CPP is broken into a number of subproblems, each of which is addressed by an AV, and the convergence of the algorithm is controlled by updating the shadow prices at the control center. Simulations reveal that the distributed algorithm can produce nearly optimal solutions with substantially reduced computational time. A coarser time scale can improve computational time but degrades the solution quality resulting in possible infeasible solution. Even with communication loss, the distributed algorithm can still perform well and converge with little degradation in speed.

\textcolor{black}{
For future work, we will extend our deterministic formulation to stochastic or robust optimization. Since some parameters, like travel times, demand profiles, and vehicle return locations, can become probabilistic in practice, formulating the problem in a probabilistic form may make this line of research more realistic.
}


%
%
%

\bibliographystyle{IEEEtran}
\bibliography{IEEEabrv}

\begin{thebibliography}{10}
\providecommand{\url}[1]{#1}
\csname url@samestyle\endcsname
\providecommand{\newblock}{\relax}
\providecommand{\bibinfo}[2]{#2}
\providecommand{\BIBentrySTDinterwordspacing}{\spaceskip=0pt\relax}
\providecommand{\BIBentryALTinterwordstretchfactor}{4}
\providecommand{\BIBentryALTinterwordspacing}{\spaceskip=\fontdimen2\font plus
\BIBentryALTinterwordstretchfactor\fontdimen3\font minus
  \fontdimen4\font\relax}
\providecommand{\BIBforeignlanguage}[2]{{%
\expandafter\ifx\csname l@#1\endcsname\relax
\typeout{** WARNING: IEEEtran.bst: No hyphenation pattern has been}%
\typeout{** loaded for the language `#1'. Using the pattern for}%
\typeout{** the default language instead.}%
\else
\language=\csname l@#1\endcsname
\fi
#2}}
\providecommand{\BIBdecl}{\relax}
\BIBdecl

\bibitem{smartgridcomm2016}
A.~Y.~S. Lam, J.~J.~Q. Yu, Y.~Hou, and V.~O.~K. Li, ``Coordinated autonomous
  vehicle parking for vehicle-to-grid services,'' in \emph{Proc. 7th IEEE Int.
  Conf. on Smart Grid Commun.}, Sydney, Australia, 2016.

\bibitem{V2G_capacity}
A.~Y.~S. Lam, K.-C. Leung, and V.~O.~K. Li, ``Capacity estimation for
  vehicle-to-grid frequency regulation services with smart charging
  mechanism,'' \emph{IEEE Trans. Smart Grid}, vol.~7, no.~1, pp. 156--166, Jan.
  2016.

\bibitem{EVmobility}
D.~Dallinger, D.~Krampe, and M.~Wietschel, ``Vehicle-to-grid regulation
  reserves based on a dynamic simulation of mobility behavior,'' \emph{IEEE
  Trans. Smart Grid}, vol.~2, no.~2, pp. 302--313, Jun. 2011.

\bibitem{trend1}
D.~Mohr, H.-W. Kaas, P.~Gao, D.~Wee, and T.~M\"{o}ller, ``Automotive revolution
  - persepective towards 2030: How the convergence of disruptive
  technology-driven trends could transform the auto industry,'' McKinsey \&
  Company, Tech. Rep., Jan. 2016.

\bibitem{trend2}
T.~Litman, ``Autonomous vehicle implementation predictions implications for
  transport planning,'' Victoria Transport Policy Institute, Tech. Rep., Sept.
  2016.

\bibitem{trend3}
M.~Westervelt, E.~Han, D.~Gopalakrishna, and J.~Klion, ``Emerging technology
  trends in transportation,'' Eno Center for Transportation and ICF
  International, Tech. Rep., Feb. 2016.

\bibitem{V2G_DR}
F.~Rassaei, W.-S. Soh, and K.-C. Chua, ``Demand response for residential
  electric vehicles with random usage patterns in smart grids,'' \emph{IEEE
  Trans. Sustainable Energy}, vol.~6, no.~4, pp. 1367--1376, Oct. 2015.

\bibitem{V2G_DR2}
R.~Yu, W.~Zhong, S.~Xie, C.~Yuen, S.~Gjessing, and Y.~Zhang, ``Balancing power
  demand through {EV} mobility in vehicle-to-grid mobile energy networks,''
  \emph{{IEEE} Trans. Ind. Informat.}, vol.~12, no.~1, pp. 79--90, Feb. 2016.

\bibitem{V2G_ancillary2}
E.~Sortomme and M.~A. El-Sharkawi, ``Optimal scheduling of vehicle-to-grid
  energy and ancillary services,'' \emph{IEEE Trans. Smart Grid}, vol.~3,
  no.~1, pp. 351--359, Mar. 2012.

\bibitem{V2G_ancillary}
------, ``Optimal combined bidding of vehicle-to-grid ancillary services,''
  \emph{IEEE Trans. Smart Grid}, vol.~3, no.~1, pp. 70--79, Mar. 2012.

\bibitem{V2G_frequencyregulation}
S.~Han, S.~Han, and K.~Sezaki, ``Development of an optimal vehicle-to-grid
  aggregator for frequency regulation,'' \emph{IEEE Trans. Smart Grid}, vol.~1,
  no.~1, pp. 65--72, Jun. 2010.

\bibitem{V2GService1}
Z.~Yang, R.~Wu, J.~Yang, K.~Long, and P.~You, ``Economical operation of
  microgrid with various devices via distributed optimization,'' \emph{IEEE
  Trans. Smart Grid}, vol.~7, no.~2, pp. 857--867, Mar. 2016.

\bibitem{V2GService2}
E.~L. Karfopoulos and N.~D. Hatziargyriou, ``Distributed coordination of
  electric vehicles providing v2g services,'' \emph{{IEEE} Trans. Power Syst.},
  vol.~31, no.~1, pp. 329--338, Jan. 2016.

\bibitem{AV_obstacle}
G.~Franze and W.~Lucia, ``A receding horizon control strategy for autonomous
  vehicles in dynamic environments,'' \emph{{IEEE} Trans. Control Syst.
  Technol.}, vol.~24, no.~2, pp. 695–--702, Mar. 2016.

\bibitem{AV_path}
C.~Chen, Y.~Jia, M.~Shu, and Y.~Wang, ``Hierarchical adaptive path-tracking
  control for autonomous vehicles,'' \emph{{IEEE} Trans. Intell. Transp.
  Syst.}, vol.~16, no.~5, pp. 2900–--2912, Oct. 2015.

\bibitem{kinect}
J.~Hernández-Aceituno, R.~Arnay, J.~Toledo, and L.~Acosta, ``Using kinect on
  an autonomous vehicle for outdoors obstacle detection,'' \emph{IEEE Sensors
  J.}, vol.~16, no.~10, pp. 3603--3610, May 2016.

\bibitem{AV_flocking}
T.-T. Han and S.~S. Ge, ``Styled-velocity flocking of autonomous vehicles: A
  systematic design,'' \emph{{IEEE} Trans. Autom. Control}, vol.~60, no.~8, pp.
  2015–--2030, Aug. 2015.

\bibitem{AVPTS}
A.~Y.~S. Lam, Y.-W. Leung, and X.~Chu, ``Autonomous vehicle public
  transportation system: Scheduling and admission control,'' \emph{{IEEE}
  Trans. Intell. Transp. Syst.}, vol.~17, no.~5, pp. 1210–--1226, May 2016.

\bibitem{AVPTS_auction}
A.~Y.~S. Lam, ``Combinatorial auction-based pricing for multi-tenant autonomous
  vehicle public transportation system,'' \emph{{IEEE} Trans. Intell. Transp.
  Syst.}, vol.~17, no.~3, pp. 859–--869, Mar. 2016.

\bibitem{connectedAV}
E.~Uhlemann, ``Autonomous vehicles are connecting...'' \emph{{IEEE} Veh.
  Technol. Mag.}, vol.~10, no.~2, pp. 22--25, Jun. 2015.

\bibitem{googlecar}
\BIBentryALTinterwordspacing
L.~Gannes. (2014, May) Google’s new self-driving car ditches the steering
  wheel. [Online]. Available:
  \url{http://recode.net/2014/05/27/googles-new-self-driving-car-ditches-the-steering-wheel/}
\BIBentrySTDinterwordspacing

\bibitem{teslacar}
\BIBentryALTinterwordspacing
R.~Bradley. (2014, Oct.) The electric-vehicle maker sent its cars a software
  update that suddenly made autonomous driving a reality. [Online]. Available:
  \url{https://www.technologyreview.com/s/600772/10-breakthrough-technologies-2016-tesla-autopilot/}
\BIBentrySTDinterwordspacing

\bibitem{parkingassistance}
E.~Kokolaki, M.~Karaliopoulos, and I.~Stavrakakis, ``Leveraging information in
  parking assistance systems,'' \emph{{IEEE} Trans. Intell. Transp. Syst.},
  vol.~16, no.~5, pp. 2913--2924, Oct. 2015.

\bibitem{streetparking}
T.~Rajabioun and P.~A. Ioannou, ``On-street and off-street parking availability
  prediction using multivariate spatiotemporal models,'' \emph{{IEEE} Trans.
  Veh. Technol.}, vol.~62, no.~9, pp. 4309--4317, Nov. 2013.

\bibitem{valet1}
K.~Min and J.~Choi, ``A control system for autonomous vehicle valet parking,''
  in \emph{Proc. 13th Int. Conf. Control, Automation and Syst.}, Gwangju,
  Korea, 2013.

\bibitem{valet2}
K.-W. Min and J.-D. Choi, ``Design and implementation of an intelligent vehicle
  system for autonomous valet parking service,'' in \emph{Proc. 10th Asian
  Control Conf.}, Kota Kinabalu, Malaysia, 2015.

\bibitem{parkingmanagement}
R.~E. Barone, T.~Giuffre, S.~M. Siniscalchi, M.~A. Morgano, and G.~Tesoriere,
  ``Architecture for parking management in smart cities,'' \emph{IET Intell.
  Transp. Syst.}, vol.~8, no.~5, pp. 445--452, 2014.

\bibitem{grid_parking}
S.~Rezaee, E.~Farjah, and B.~Khorramdel, ``Probabilistic analysis of plug-in
  electric vehicles impact on electrical grid through homes and parking lots,''
  \emph{IEEE Trans. Sustainable Energy}, vol.~4, no.~4, pp. 1024--1033, Oct.
  2013.

\bibitem{smartcity}
S.~P. Mohanty, U.~Choppali, and E.~Kougianos, ``Everything you wanted to know
  about smart cities: The internet of things is the backbone,'' \emph{{IEEE}
  Consum. Electron. Mag.}, vol.~5, no.~3, pp. 60--70, Jul. 2016.

\bibitem{dijkstra}
T.~H. Cormen, C.~E. Leiserson, R.~L. Rivest, and C.~Stein, \emph{Introduction
  to Algorithms}, 2nd~ed.\hskip 1em plus 0.5em minus 0.4em\relax Cambridge, MA:
  MIT Press, 2001.

\bibitem{V2GApp1}
E.~Sortomme and M.~A. El-Sharkawi, ``Optimal charging strategies for
  unidirectional vehicle-to-grid,'' \emph{{IEEE} Trans. Smart Grid}, vol.~2,
  no.~1, pp. 131--138, Mar. 2011.

\bibitem{V2GApp2}
M.~Singh, K.~Thirugnanam, P.~Kumar, and I.~Kar, ``Real-time coordination of
  electric vehicles to support the grid at the distribution substation level,''
  \emph{{IEEE} Syst. J.}, vol.~9, no.~3, pp. 1000--1010, Sept. 2015.

\bibitem{V2GApp3}
M.~J.~E. Alam, K.~M. Muttaqi, and D.~Sutanto, ``Effective utilization of
  available {PEV} battery capacity for mitigation of solar {PV} impact and grid
  support with integrated {V2G} functionality,'' \emph{{IEEE} Trans. Smart
  Grid}, vol.~7, no.~3, pp. 1562--1571, May 2016.

\bibitem{V2GApp4}
K.~Kaur, R.~Rana, N.~Kumar, M.~Singh, and S.~Mishra, ``A colored petri net
  based frequency support scheme using fleet of electric vehicles in smart grid
  environment,'' \emph{{IEEE} Trans. Power Syst.}, vol.~31, no.~6, pp.
  4638--4649, Nov. 2016.

\bibitem{VANET}
D.~N. Cottingham, ``Vehicular wireless communication,'' University of
  Cambridge, Tech. Rep. UCAM-CL-TR-741, Jan. 2009.

\bibitem{bertsekas}
D.~P. Bertsekas, \emph{Nonlinear Programming}.\hskip 1em plus 0.5em minus
  0.4em\relax Belmont, MA: Athena Scientific, 1999.

\bibitem{cdc}
A.~Y.~S. Lam, B.~Zhang, and D.~Tse, ``Distributed algorithms for optimal power
  flow problem,'' in \emph{Proc. 51st IEEE Conf. on Decision and Control},
  Maui, HI, 2012.

\bibitem{voltage}
B.~Zhang, A.~Y.~S. Lam, A.~Dominguez-Garcia, and D.~Tse, ``An optimal and
  distributed method for voltage regulation in power distribution systems,''
  \emph{{IEEE} Trans. Power Syst.}, vol.~30, no.~4, pp. 1714–--1726, Jun.
  2015.

\bibitem{cvx}
S.~Boyd and L.~Vandenberghe, \emph{Convex Optimization}.\hskip 1em plus 0.5em
  minus 0.4em\relax Cambridge, NY: Cambridge University Press, 2004.

\bibitem{gurobi-ref}
\BIBentryALTinterwordspacing
Gurobi optimization. [Online]. Available: \url{http://www.gurobi.com/}
\BIBentrySTDinterwordspacing

\bibitem{latency}
\BIBentryALTinterwordspacing
M.~Dano. (2016, Nov.) {3G/4G} wireless network latency: Comparing {Verizon},
  {AT}\&{T}, {Sprint} and {T-Mobile} in {February} 2014. [Online]. Available:
  \url{http://www.fiercewireless.com/special-report/3g-4g-wireless-network-latency-comparing-verizon-at-t-sprint-and-t-mobile-february}
\BIBentrySTDinterwordspacing

\end{thebibliography}

\end{document}